\documentstyle[aps]{revtex}
\begin{document}
\author{V. N. Bogomolov}
\title{Metallic xenon. Polarizability. Equation of state.}
\date{\today}
\address{Russian Academy of Sciences \\
A.F. Ioffe Physical \& Technical Institute\\
St. Petersburg, Russia}
\maketitle

\begin{abstract}
It is shown that some of the physical properties of inert gas ( IG )
condensates (polarizability, compressibility, metallization under pressure,
equation of state) may be well described taking into account the first
excited state of atoms only. The Herzfeld criterion of metallization well
corresponds to the Mott transition criterion and to the percolation
threshold. For metallic xenon concentration of the molecular type
excitations corresponds to the Bose condensation temperature $T_C\sim 4000K$.
The BCS formula gives $T_c\sim 5000K$. If phonons are changed by
fluctuations of interatomic interaction energy. A simple relations
between the parameters of atoms at metallization has been found.
\end{abstract}

\bigskip

One of the peculiarities of inert gas (IG) condensates is the coincidence of
interatomic distances with diameters $2r_2$ of atoms in the first excited
state. This is the reason to ignore the next excited states. Such atoms are
excimer analogies of the corresponding alcaly metals. For this case an
alternative approach to a description ofproperties of molecular condensates
is possible \cite{bvn99}. These substances have a bond of a metallic type
via excited state orbitals but without a metallic conductivity because a
mean number of electrons at excited state orbitals $X<1$. For xenon $X=0,038$%
. Most of the physical properties ( condensation and adsorption energy,
compressibility, metallization under pressure ) may be described by
expressions of the theory of simple metals with electron charge e replaced
by $eX$ (and with replaced numeral coefficients) \cite{bvn95a,bvn93,bvn95b}.

Properties of the condensate are determined by the excited state radius $r_2$
of atoms. It may be expressed through energy by the hydrogen-like formula $%
2r_2=.e^2/(E_1-E_2)$. Here $E_1=e^2/2r_1$--- ionization potential, $E_2$ -
the transition energy between the ground and excited state of atom. For
system in which the excited state radius of atoms is fixed the atomic wave
functions may be expressed by the energy. Probability $X$ for an electron to
appear at excited state orbital with radius $r_2$ may be obtained from the
atomic wave functions of excited state:

\[
X(r_2)=X_0\exp \left[ -\frac{2E_2\left( r_2-r_1\right) }{e^2}\right]
=X_{01}\exp \left[ -\left( \frac{r_2}{r_1}+\frac{r_1}{r_2}\right) \right]
=X_{02}\exp \left( -\frac{r_2}{r_1}\right) \;\text{at\ }r_2\gg r_1\,.
\]

The pre-exponential coefficient is a relatively weak function of $r$. An
average perturbation energy $w=e^2/2\left( r_2-r_1\right) $ is the
interaction energy between electrons of the ground state orbitals of the
neighbouring atoms at distance $2r_2$ in condensate. The mean electron
concentration $X$ per atom depends on the perturbation energy $w$. The
pressure is more significant parameter then the temperature. For
ordinary semiconductors with well-defined band structure this energy
is $kT$. For IG the relation is:
$g=w/E_2\sim (0.30-0.55)$. Metallization of xenon occurs at $g_m=0,75$. The
main properties of condensates may be expressed through atomic spectroscopic
parameters of atoms. Radii of ground and excited states of atoms $r_1$ and $%
r_2$ may be changed by the corresponding atomic volumes $V_1$ and $V_2$:
\[
X(r_2)\sim \exp [-(y^{1/3}+y^{-1/3})];\;y=\frac{V_2}{V_1}=\frac{E_1^3}{%
(E_1-E_2)^3}\,.
\]
For IG volumes $V_1$ are approximately independent on pressure:

\begin{center}
\begin{tabular}{p{3cm}p{3cm}p{3cm}p{3cm}p{3cm}}
& Ne & Ar & Kr & Xe \\
$V_1(A^3)$ & 0.156 & 0.400 & 0.569 & 0.873
\end{tabular}
\end{center}

A simple physical picture of the model discussed may be revealed if applied
to calculation of the polarizability of atoms in condensates.

Atomic polarizability a well corresponds to polarizability of the excited
state orbital with the occupation number $X$:
\[
a_c\sim eXr_2\frac{er_2}{E_1-E_2}\sim Xr_2^3\sim XV_2\sim (E_1-E_2)^{-3}\exp
\left[ -\frac{E_2^2}{E_1\left( E_1-E_2\right) }\right] \,.
\]
The molar refractivity $R$ is proportional to polarizability $a$. Hence the
Herzfeld criterion of metallization \cite{hkf27} must be proportional to $%
XV_2$. We have:

\begin{center}
\begin{tabular}{p{3cm}p{3cm}p{3cm}p{3cm}p{3cm}}
& Ne & Ar & Kr & Xe \\
$R(cm^3/mol)$ & 1.00 & 4.15 & 6.25 & 10.2 \\
$XV_2(cm^3/mol)$ & 1.57 & 4.49 & 6.60 & 10.2
\end{tabular}
\end{center}

From the other side, metallization occurs at $X_p=0,12$ ($y_m=14.4$) as a
transition over the percolation threshold \cite{bvn99,bvn95a}. Consequently
the molar volume at metallization $V_{02m}\sim V_{2m}=V_1y_m=V_114.4\sim
E_1^{-3}$ for all IG . This relation well corresponds to criterion of the
Mott transition too ($V_{2m}/V_1\sim 15$).

\begin{center}
\begin{tabular}{p{3cm}p{3cm}p{3cm}p{3cm}p{3cm}}
$V_{02m}(cm^3/mol)$ & 1.73 & 4.65 & 6.63 & 10.20
\end{tabular}
\end{center}

\noindent Quantum mechanics calculations gives \cite{tbp93}:

\begin{center}
\begin{tabular}{p{3cm}p{3cm}p{3cm}p{3cm}p{3cm}}
$V_{02m}(cm^3/mol)$ & --- & 4.61 & 6.67 & 10.23
\end{tabular}
\end{center}

Metallic xenon contain virtual excimer Xe$_2$ molecules in metal type
lattice. The mean concentration of Xe$_2$ is about $0.12\times 3.015\;10^{22}$
and the Bose condensation temperature $T_C\sim 4000\,K$ (at  $m^{*}=2m_0$).
By analogy with the BCS formula $T_c\sim 0.5w_mX_p\sim 0.5E_2X_p\sim 5000K$
if phonons are changed by fluctuations of ineratomic interaction energy $m$.
It does not contradict with \cite{bvn99}. For the hypothetical metallic H$_2$
(H$_4$ virtual molecules), $T_C\sim 10000\,K$ at $V_{02m}\sim 3\,cm^3/mol$.

The next verification of the model may be made if used to calculation of the
compressibility and equation of state (EOS). Metal-like compressibility:
\[
-\frac{dV_2}{V_2dP_c}\sim \frac{r_2^4}{e^2X^2}\sim \frac{V^{4/3}y^{4/3}}{%
\exp \left[ -2\left( y^{1/3}+y^{-1/3}\right) \right] }
\]
EOS at $T=0$ and $y>60$ is:
\begin{eqnarray*}
P_c(V_0) &=&P_0V_1^{-4/3}\int \left( y-l\right) ^{-7/3}\exp \left\{ -2\left[
\left( y-l\right) ^{1/3}+\left( y-l\right) ^{-1/3}\right] \right\} dy \\
&=&P_{10}E_1^4F\left( E_1;V_0\right) +P_{20}=P_{30}F\left( E_1;V_0\right)
+P_{20}\,,
\end{eqnarray*}
\begin{eqnarray*}
P_{10} &=&173.74827(GPa;\,eV^{-7});\;P_{20}=-1.294312(GPa); \\
P_{30} &=&3.7578024\,10^6(GPa)\;\text{at }E_1=12.127\,eV\;\text{for xenon.}
\end{eqnarray*}

The volume $V_1$ of atom in ground state is subtracted from the volume $V_2$
(i.e. parameter $y$ replaced by $y-1=(V_2-V_1)/V_1$.

The metallization pressure is $P_m\sim E_1^4$ for molecular condensates.

The $P_e$ vs. $V_0$ data \cite{gesm89} are presented below together with the
calculated $P_c$ for xenon:

\begin{center}
\begin{tabular}{p{2.5cm}p{1.2cm}p{1.2cm}p{1.2cm}p{1.2cm}p{1.2cm}p{1.2cm}%
p{1.2cm}p{1.2cm}p{1.2cm}p{1.2cm}p{1.2cm}}
$V_0(cm^3/mol)$ & 34.7 & 11.41 & 11.1 & 10.88 & 10.75 & 10.46 & 10.24 & 10.12
& 9.88 & 9.72 & 9.4 \\
$P_e(GPa)$ & 1 10$^{-4}$ & 107 & 117 & 125 & 130 & 142 & 152 & 158 & 171 &
180 & 200 \\
$P_c(GPa)$ & 1 10$^{-4}$ & 107.5 & 118 & 125.5 & 130 & 142 & 151.5 & 157 &
169.5 & 178 & 197.5 \\
$y$ & 48.80 & 16.1 & 15.6 & 15.3 & 15.1 & 14.7 & 14.4 & 14.2 & 13.9 & 13.7 &
13.2
\end{tabular}
\end{center}

Again, the agreement is good. The main purpose of these evaluations is to
find extremely simple and adequate description of the intermolecular and
metal-dielectric interactions alternative to traditional way. It may be
useful for understanding of some problems of nanoelectronics and of
multicomponent HTSC.

\end{document}